\documentclass[prb,reprint,amsmath,amssymb,showpacs,superscriptaddress,floatfix]{revtex4-1}
\usepackage[breaklinks=true,colorlinks,citecolor=blue,linkcolor=blue,urlcolor=blue]{hyperref}
\usepackage{epsfig,mathrsfs,color,latexsym,subfigure,marginnote,gensymb,}
\usepackage{graphicx}
\usepackage{xcolor}
\usepackage{chemformula}
\usepackage{tabularx}
\usepackage{makecell}
\renewcommand{\BibitemShut}[1]{}

\usepackage[utf8]{inputenc}

\begin{document}
\title{Versatility of type-II van der Waals heterostructures: a case study with SiH-CdCl$_2$}

\author{Achintya Priydarshi}
\affiliation{Department of Electrical Engineering, Indian Institute of Technology Kanpur, Kanpur 208016, India}
\author{Abhinav Arora}
\affiliation{Department of Materials Science and Engineering, Indian Institute of Technology Kanpur, Kanpur 208016, India}
\author{Yogesh Singh Chauhan}
\email[]{chauhan@iitk.ac.in}
\affiliation{Department of Electrical Engineering, Indian Institute of Technology Kanpur, Kanpur 208016, India}
\author{Amit Agarwal}
\email[]{amitag@iitk.ac.in}
\affiliation{Department of Physics, Indian Institute of Technology Kanpur, Kanpur 208016, India}
\author{Somnath Bhowmick}
\email[]{bsomnath@iitk.ac.in}
\affiliation{Department of Materials Science and Engineering, Indian Institute of Technology Kanpur, Kanpur 208016, India}

\date{\today}
\begin{abstract}
Unlike bilayers or a few layers thick materials, heterostructures are designer materials formed by assembling different monolayers in any desired sequence. As a result, 
heterostructures can be tailor-made for specific functionalities and applications. Here, we show the potential of heterostructures for several applications, like piezoelectricity, photocatalytic water splitting, and tunnel field effect transistor (TFET), by taking SiH-CdCl$_2$ as a representative system. Our study reveals that the characteristics of the heterostructure mainly depend on the potential difference between the constituent monolayers. By leveraging the extensive database of layered materials, many such combinations with a suitable potential difference can be formed 
for specific functionalities. 
This opens up exciting avenues for designing and engineering materials with tailored properties, paving the way for next-generation device applications. 
\end{abstract}
\maketitle
\section{INTRODUCTION}
Since the experimental discovery of graphene \cite{Novoselov666}, numerous two-dimensional (2D) materials have been extensively explored due to their exceptional electrical, mechanical, optical, and catalytic properties. These include transition metal dichalcogenides, group-V materials like phosphorene, and group-IV materials like silicene among others \cite{chhowalla2013chemistry,zhang2017two,doi:10.1021/nn501226z, doi:10.1021/jp508618t,nahaspccp,sougataprb, DI201779,silecenepss}. To impart additional functionality to 2D materials beyond their intrinsic properties, various techniques such as chemical functionalization, surface adsorption, strain engineering, and quantum confinement are used.  \cite{graphaneprb,priyankjpc,nahasprb,puriprb,achintyaprb,somnathprb}. Owing to their tunable properties, van der Waals (vdW) heterostructures have emerged as a promising class of 2D materials \cite{geim2013van,Novoselovaac9439}. These heterostructures are are formed by stacking monolayers in desired sequences, allowing for the tailoring of material properties for specific applications \cite{deng2016catalysis,bertolazzi2013nonvolatile}.

Heterostructures can be broadly classified into three types based on their electronic properties: type I, type II, and type III. Among these, type II heterostructures exhibit distinct advantages due to the presence of valence and conduction band edges localized at different monolayers \cite{heterotype}. This band alignment is particularly advantageous for applications such as solar energy harvesting and photocatalytic water splitting. The latter uses solar energy to generate oxygen and hydrogen, an excellent way to produce clean energy in the future. For both these applications, efficient charge separation is essential to minimize electron-hole recombination.  
Owing to this, Type-II heterostructures have been proposed to be potential candidates for these applications \cite{rawat2019solar,mao2019two,wang2013visible,he2019type,li2021two,MAO2023156298}. Additionally, type II heterostructures have shown promise in other areas, including tunneling field effect transistors (TFETs) \cite{zhu2013band} and piezoelectricity \cite{MOHANTA2021150928}.

Although previous studies have highlighted various applications of van der Waals heterostructures individually, a systematic exploration of their multifunctional aspects is lacking. In this work, we investigate the properties of the SiH-CdCl$_2$ heterostructure using \textit{ab initio} calculations. While the individual monolayers, namely silicane (SiH) and CdCl$_2$, have been studied separately, their combination as a heterostructure and the resulting physical properties are relatively unexplored \cite{zhang2012first, qiu2015silicene, houssa2011electronic,restrepo2014tunable, sheng2020inse, han2020ptse, luo2019highly}. We show that SiH-CdCl$_2$ heterostructure exhibits a type-II band alignment, making it a potential candidate for efficient photocatalytic water splitting applications. To confirm its potential as photocatalyst utilizing solar energy, we calculate its absorption spectrum and show that its absorption peaks lie in the visible range.  
Additionally, we explore the application of the heterostructure in TFETs, and peizoeletricity. Our findings reveal that the characteristics of the heterostructure depend on the potential difference.

Our paper is organized in the following manner: Section~\ref{methodology} lists the details of the \textit{ab initio} calculations. In Section~\ref{rd}, we present the results, starting with crystal structure in \ref{rd}A, followed by electronic and optical properties in \ref{rd}B and \ref{rd}C, respectively. We illustrate the charge distribution at the heterojunction in \ref{rd}D, and discuss the resulting piezoelectricity is discussed in \ref{PZE}. We describe other applications like photocatalysis and TFET in Sections \ref{PhotoCt} and \ref{TFET}, respectively. We conclude the paper in Section~\ref{conclusion}.

\section{Methodology}
\label{methodology}
For our \textit{ab initio} calculations, we used the density functional theory (DFT) and projector-augmented wave pseudo-potentials, as implemented in the Quantum Espresso package \cite{Giannozzi_2009}. Exchange and correlation energies are calculated using the Perdew-Burke-Ernzerhof (PBE) implementation of the generalized gradient approximation (GGA). We set the energy cutoff for the plane wave basis to be 60 Ry. A $k$-mesh of  $14 \times14\times 1$ is used for the 2D Brillouin zone (BZ) integrations. For structural relaxations, we use force and energy convergence threshold of $10^{-3}$ Ry/a.u and $10^{-4}$ Ry, respectively. We use Grimme's DFT-D method to capture the effects of the vdW interactions \cite{grimme2006semiempirical}. To eliminate the spurious interactions between neighboring slabs, a vacuum layer of 20~\AA~is inserted between adjacent layers along the $z-$direction (perpendicular to the plane of monolayers). 
VESTA is used for visualizing crystal structures  \cite{momma2008vesta}. We use the Heyd–Scuseria–Ernzerhof (HSE06) hybrid functional to correct for the known bandgap underestimation within the standard semilocal DFT approaches\cite{hse}. 

\begin{figure}    
\includegraphics[width=1.0\linewidth]{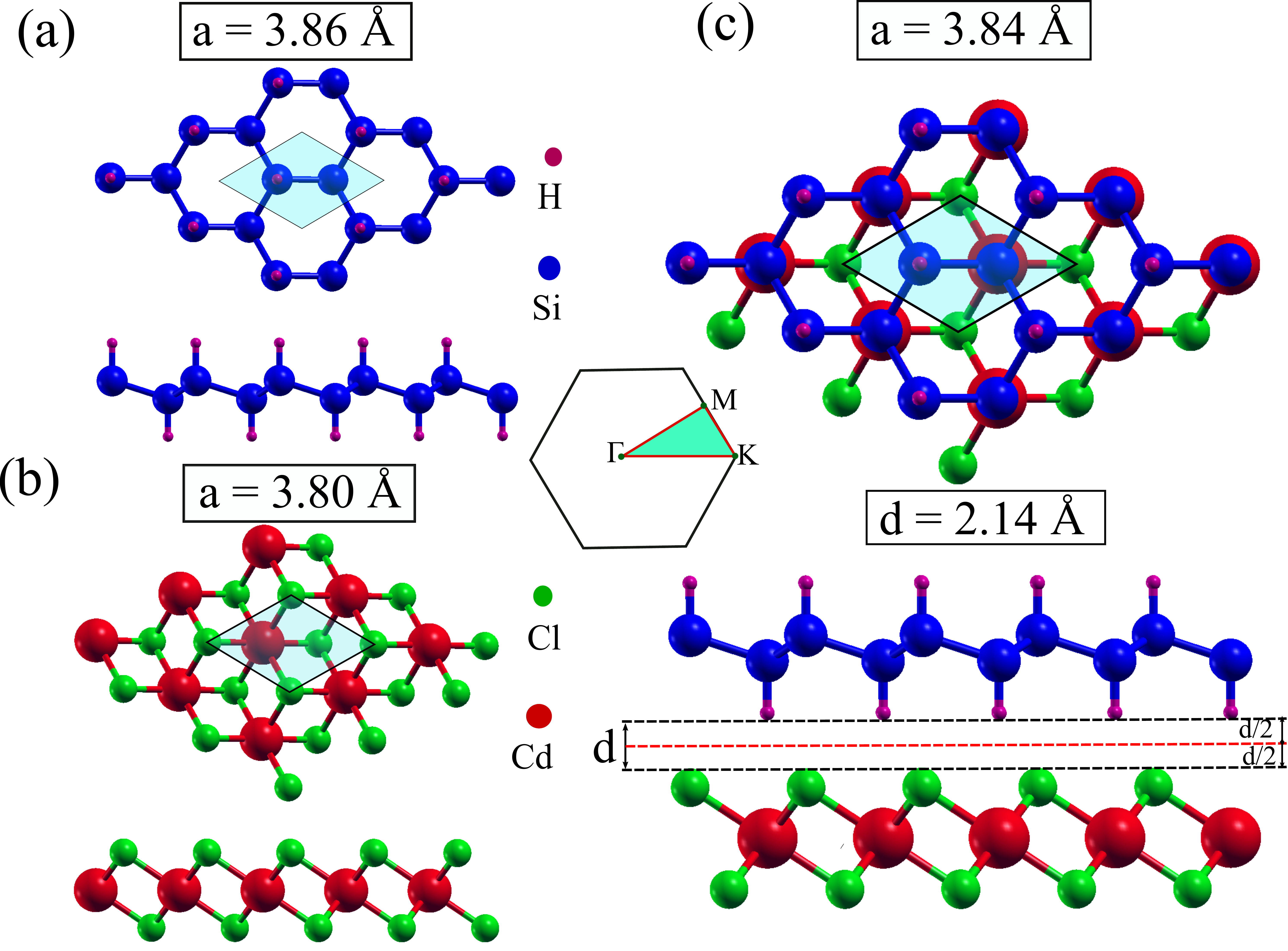}
\caption{ Top view (top panel) and side view (bottom panel) of isolated monolayers of (a) SiH, (b) CdCl$_{2}$, and (c) the most stable stacking among the six possible stacking of their heterostructure. We define the interface as located at a distance of $d/2$ from the SiH and CdCl$_2$ monolayers.}
\label{fig:structure}
\end{figure}

\begin{table}
\caption{Calculated lattice parameter ($a$), band gap (E$_g$), and band edge location ($E_{\rm CBM}$ and $E_{\rm VBM}$) for isolated SiH, CdCl$_{2}$ monolayers and their heterostructure. $E_{\rm CBM}$ and $E_{\rm VBM}$ are determined with reference to the vacuum level. The value of E$_g$ within the brackets and outside corresponds to PBE and HSE-based calculations, respectively.\label{T1}}
\centering
{
\begin{tabular}{c c c c c c c c}
\hline
Structure    & $a$~(\AA) & $E_{\rm CBM}$ (eV) & $E_{\rm VBM}$ (eV) & $E_{g}$  (eV)\\
\hline
SiH  & 3.86 & -2.59 & -5.54 & 2.95 (2.18) &\\
CdCl$_{2}$ & 3.80 & -3.39 &-8.73 & 5.34 (3.98) &\\
SiH/CdCl$_{2}$ & 3.84 & -3.34 & -5.77 & 2.43 (1.18) & \\
\hline
\end{tabular} }\\
\end{table}

\section{Physical properties}
\label{rd}
\subsection{Crystal structure}
Both SiH and CdCl$_{2}$ monolayers have a hexagonal unit cell [see Figure~\ref{fig:structure}(a) and (b)], and the optimized lattice parameters are found to be 3.86~\AA \ and 3.80~\AA, respectively. Other structural parameters for both the monolayers we obtained [see Table S1 of Supporting Information (SI) for further details] also agree with the reported values \cite{han2020ptse,han2021alas,yue2021single,kistanov2022family}. This validates our choice of computational parameters reported in Section~\ref{methodology}. Out of several possible stacking sequences of monolayer SiH and CdCl$_2$ [see Figure S1 in SI for further details], the one with the lowest energy is shown in Figure~\ref{fig:structure}(c). The most stable structure has a lattice parameter of 3.83~\AA. The parent monolayers and the heterostructure have very similar lattice parameters, with the difference being less than 1\%. This ensures that none of the layers are under significant strain in the heterostructure. The interlayer separation in this heterostructure (2.14~\AA) is significantly less than that of bilayer graphene (3.4~\AA). This indicates some electrostatic interaction (in addition to the Van der Waals interaction) among the parent monolayers of the heterostructure supported by charge redistribution at the interface.

We calculate the binding energy between the monolayers using \cite{almayyali2020stacking}
\begin{equation}
E_b = (E_{SiH/CdCl_{2}} - E_{SiH} - E_{CdCl_{2}})/S_{0},
\end{equation}
where E$_{SiH/CdCl_{2}}$, E$_{SiH}$, and E$_{CdCl_{2}}$ are the total energy of the optimized heterostructure and monolayers, respectively. $S_0$ is the area of the optimized heterostructure. We find the binding energy of the most stable stacking to be -7.44 meV/\AA$^2$, which is comparable to that of other vdW heterostructures like InSe/SiH (-21.96 meV/\AA$^2$) and g-C$_3$N$_4$/MoS$_2$ (-17.8 meV/\AA$^2$) \cite{sheng2020inse, wang2014enhanced}. A negative binding energy suggests sufficient binding between the two constituent monolayers of the heterostructure.

\begin{figure*}
\includegraphics[width=1.0\linewidth]{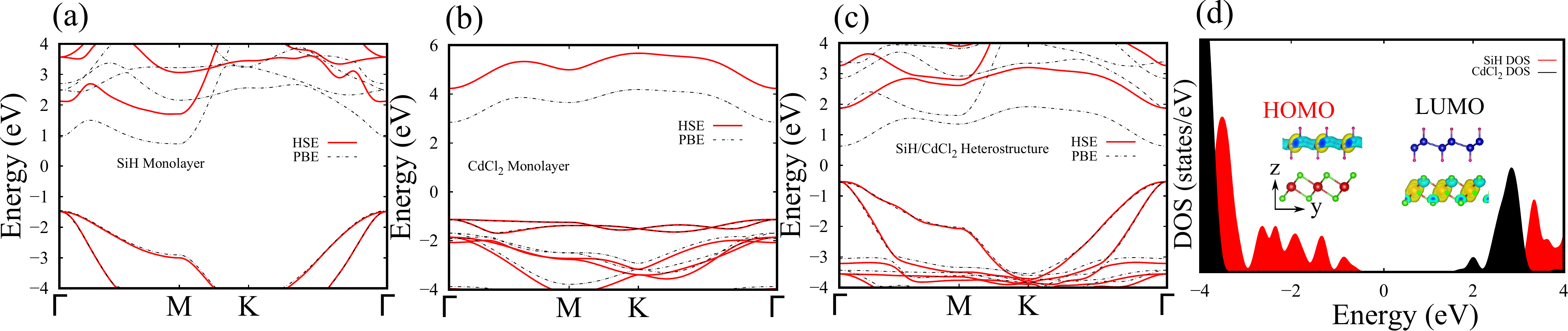}
\caption{Band structures for (a) SiH monolayer, (b) CdCl$_2$ monolayer, and (c) SiH/CdCl$_2$ heterostructure. The black (dashed) lines represent the GGA-PBE results. Red (solid) lines represent the HSE results. (d) The projected density of states of the SiH/CdCl$_{2}$ heterostructure and corresponding highest occupied molecular orbital (HOMO) and the lowest unoccupied molecular orbital (LUMO). The density of states clearly shows that SiH-CdCl$_{2}$ heterostructure has a type-II band alignment with the  HOMO and LUMO being confined entirely to SiH and CdCl$_2$ layer, respectively.}
\label{fig:BS_all}
\end{figure*}

\begin{figure}  
\includegraphics[width=0.9\linewidth]{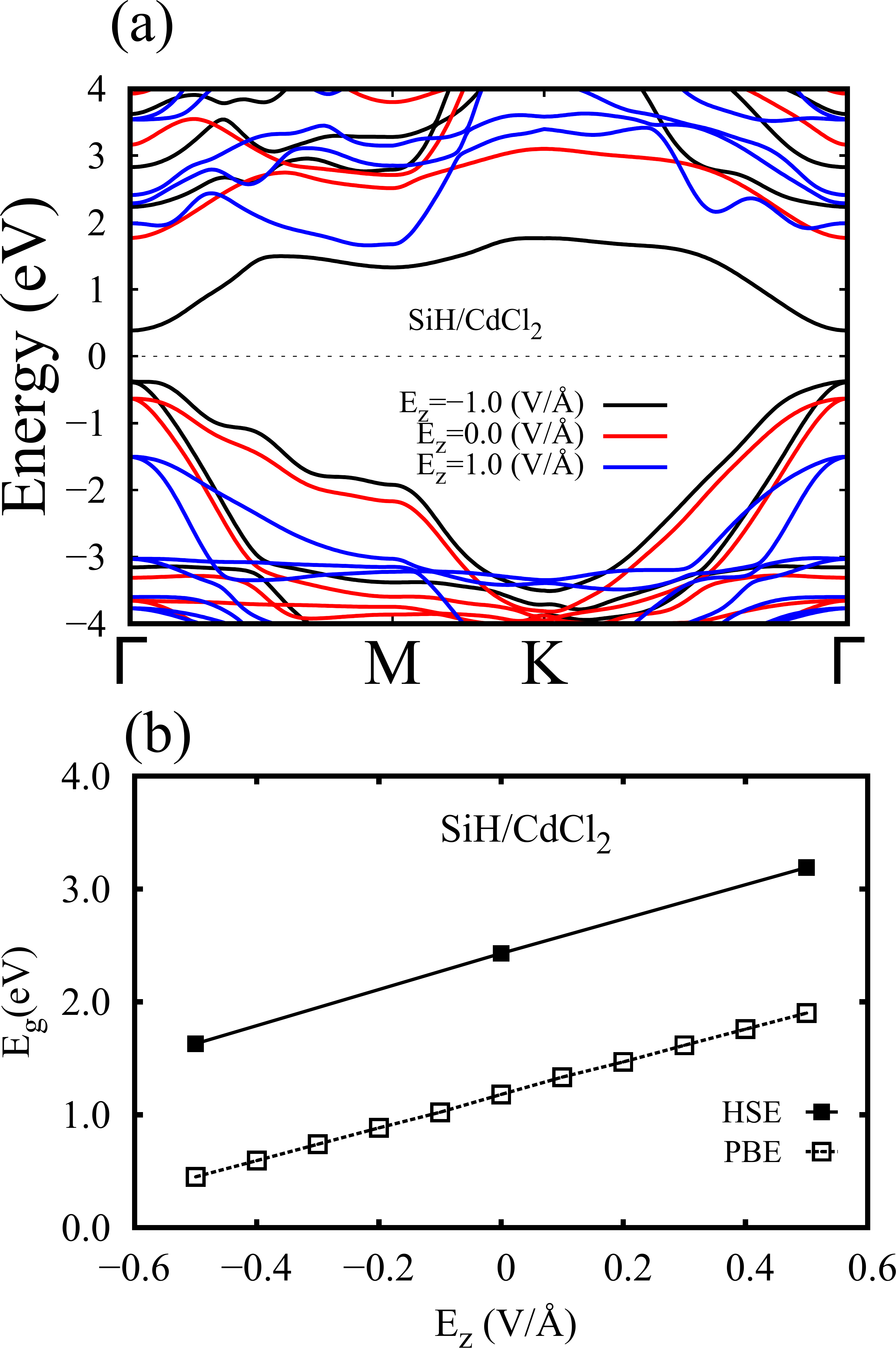}
\caption{ (a) The modification in the electronic band structure and (b) the band gap variation of SiH/CdCl$_2$ heterostructure as a function of vertical electric field E$_z$.}
\label{fig:E-Eg}
\end{figure}
\subsection{Electronic band structure and carrier mobility}
\label{secebs}
The electronic band structure is the most fundamental characteristic for determining the suitability of a material for electronic and optical applications. We perform the electronic band structure calculations using the optimized structures described in the previous section. Among several possible stackings, the one with the lowest energy is chosen for the heterostructure. The electronic band structures of isolated SiH monolayer, isolated CdCl$_{2}$ monolayer, and SiH-CdCl$_2$ heterostructure, calculated by using PBE and HSE06 methods, are shown in Figure \ref{fig:BS_all} (a), (b), and (c), respectively. SiH monolayer is a semiconductor having an indirect band gap of 2.95 eV. The valence band maximum (VBM) and conduction band minimum (CBM) is at the $\Gamma$ and M-point, respectively. CdCl$_{2}$ monolayer is a direct band gap semiconductor with a band gap of 5.34 eV. Both CBM and VBM are located at the $\Gamma-$point. Electronic band structures of SiH and CdCl$_{2}$ monolayers agree with the reported theoretical and experimental studies  \cite{han2021alas,yue2021single,kistanov2022family}. As shown in Figure \ref{fig:BS_all}(c), SiH-CdCl$_2$ heterostructure is a direct band gap semiconductor with a band gap of 2.43 eV. This is lower than the bandgap of both the parent compounds. Both CBM and VBM are located at the $\Gamma$ point. Projected density of states (PDOS), as well as HOMO-LUMO plots [integrated local density of states near the band extremum, as shown in Figure \ref{fig:BS_all}(d)] further confirm that the VBM and CBM states are confined to the SiH and CdCl$_2$ layer, respectively. Such a configuration is known as the type-II band alignment, which is very effective for electron-hole separation. We show below that the type-II band alignment makes SiH-CdCl$_2$ heterostructure a suitable candidate for photocatalysis.

Bandstructure tuning via an out-of-plane electric field is a crucial feature of layered materials. Our calculations show that an out-of-plane electric field can also engineer the electronic properties of SiH-CdCl$_2$ heterostructure [Figure \ref{fig:E-Eg}]. We find that for an electric field applied in the positive z-direction, the SiH bands move downwards, and CdCl$_2$ bands move upwards, leading to an overall bandgap increase. For a very high electric field (say 1.0 V/\AA), we see a shift from Type-II to Type-I band alignment in this heterostructure, with both VBM and CBM belonging to the SiH. Such tunability in band alignment can be exploited in applications like optoelectronic devices and thin film transistors (TFTs) \cite{GAO2020147026, Novoselovaac9439}. We find that other than the extreme case when the band structure changes from type-II to type-I, for smaller electric field values, the bandgap changes linearly [Figure \ref{fig:E-Eg} (b)] while maintaining the type-II alignment. For an electric field applied in the negative (positive) z-direction, we find that the band edges of SiH and CdCl$_2$ move close to (away from) each other to decrease (increase) the bandgap. Such a linear change can be attributed to the formation of the heterostructure, as we do not find any change in bandgap in isolated monolayers of SiH and CdCl$_2$ by varying electric field from -0.5 V/\AA~ to 0.5 V/\AA [see Figure S2 in the SI]. 
The externally applied field opposes or favors the internal field (generated by the intrinsic interlayer potential). As a result, the bandgap changes depending on the direction of the externally applied electric field. Later we shall discuss in detail how such an electric field tunable bandgap is suitable for tunnel field effect transistors.

\begin{table*}
\small
\caption{Effective mass, mobility and other related parameters of the heterostructure.}
\label{mobility}
\begin{tabular*}{\textwidth}{@{\extracolsep{\fill}}lllllllll}
\hline
Carrier & $m_x^*⁄m_0$ & $m_y^*⁄m_0$ & $C_x$ & $C_y$ & $E_x$ & $E_y$  & $\mu_x$ (zigzag) & $\mu_y$ (armchair)\\
&  &  & (J/m$^2$) & (J/m$^2$) & (eV) & (eV) & $cm^2/V-s$ & $cm^2/V-s$\\
\hline
\\
e & 1.04 & 1.04 & 175.453 & 173.925 & 5.925 & 6.711 & 96.7362 & 75.3983\\
h & 0.497 & 0.497 & 175.453 & 173.925 & 5.082 & 6.358 & 575.599 & 364.642\\
\hline
\end{tabular*}
\end{table*}

Having explored the electronic properties, we now focus on carrier mobility, an important property for transistors and optoelectronic device applications. At low energies, the electron-acoustic phonon coupling dominates the scattering process \cite{PhysRevB.87.235312}. Acoustic phonon-limited carrier mobility of the negative and positive charge carriers of SiH/CdCl$_2$ heterostructure is calculated using the deformation potential (DP) theory by Bardeen and Shockley \cite{PhysRevB.87.235312}.  The phonon-limited mobility of charge carriers in a 2D crystal along the $~\alpha$ ($\alpha = x/y$) direction is given by,
\begin{equation} \label{mobility}
	\mu_{\alpha}=\dfrac{2e\hbar^{3}C_{\alpha}}{3k_{B}Tm^{*}_{\alpha}\sqrt{m^{*}_{\alpha}m^{*}_{\beta}}(E_{\alpha}^{i})^{2}}~.
\end{equation}
Here, $m^{*}_{\alpha}$ and $m^{*}_{\beta}$ represent the effective mass of the carriers along the transport and the transverse directions, $T$ is the temperature, $C_{\alpha}$ is the longitudinal acoustic (LA) phonon's elastic modulus in the direction of transport, and $E_{\alpha}$ is the DP constant.

The in-plane elastic modulus ($C_{\alpha}$) along the $\alpha$ direction for 2D systems is given by the equation $(E-E_{0})/S_{0}=\frac{1}{2}C_{\alpha}(\delta l_{\alpha}/l_{\alpha 0})^{2}$, where $E-E_{0}$ is the change in total energy owing to applied uniaxial strain $\delta l_{\alpha}/l_{\alpha 0}$, and $S_{0}$ is the area of the pristine unit cell. $C_{\alpha}$ is calculated by fitting the strain vs. energy curve (calculated from first principles) with a parabolic function, with the strain varied between -0.5\% and 0.5\% [see Figure S3 (a) in SI]. We determine the DP constant from the strain-induced shift of the band edges (conduction band minimum for electrons and valence band maximum for holes). The DP constant is defined as $E_{\alpha}^{i}=\delta E_{i}/(\delta l_{\alpha}/l_{\alpha 0})$, where $\delta E_{i}$ is the energy change of the $i^{th}$ band (VBM and CBM for holes and electrons, respectively) under the applied strain of $\delta l_{\alpha}/l_{\alpha 0}$. A linear fit of $\delta E_{i}$ vs. strain curve (obtained from first principle calculations) is used to compute $E_{\alpha}^{i}$ with the strain varying from $-0.5$ to $0.5$ percent [see Figure S3 (b) in SI]. The effective mass $m^{*}_{\alpha} \equiv m^{*}_{\alpha \alpha}=\hbar^{2}/(\partial^2 E(k)/\partial k_\alpha \partial k_\alpha)$ of the electron (hole) is calculated using a quadratic fit of the energy dispersion in the vicinity of the minima (maxima) of the conduction (valance) band.

We calculate the carrier mobility at room temperature, i.e., $T=300$ K [Table~\ref{mobility}]. The hole mobilities are almost six times higher than that of electrons. This is mainly because of a lower hole effective mass than the electron. Our calculations reveal significant anisotropy, with both electrons and holes having 1.3 to 1.5 times higher carrier mobility along the zigzag direction. This arises from the the lower DP constant of electrons and holes in the zigzag direction than that of the armchair direction. In the next subsection, we explore the optical properties of the heterostructure. 

\subsection{Optical absorption}
\label{secopt}
\begin{figure}[t]
\centering
\includegraphics[width=0.49\textwidth]{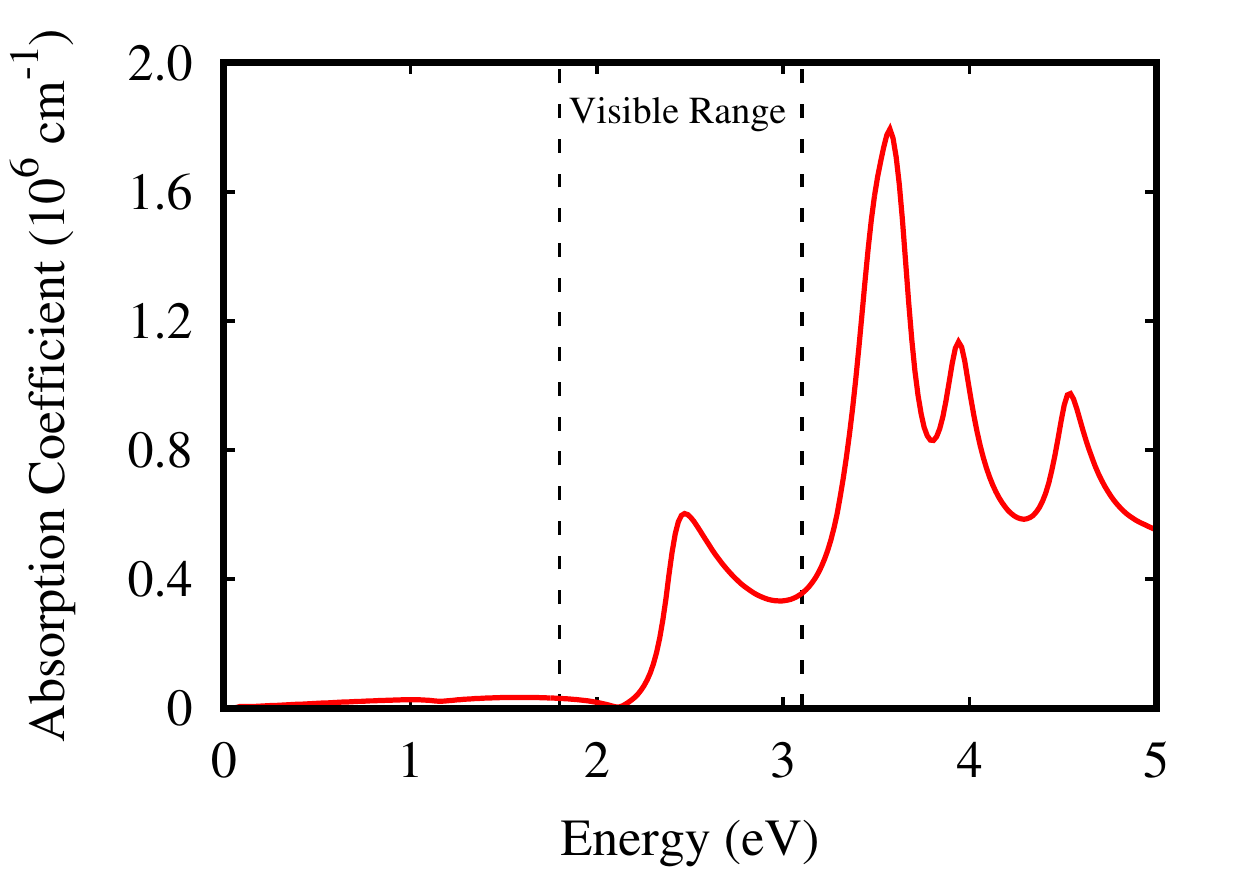}
\caption{ The variation of the total in-plane absorption coefficient with the incident photon's energy.}
\label{absorption}
\end{figure}
The optical absorption coefficient is calculated using the following expression, \cite{PhysRevB.73.045112}
\begin{equation}
	\alpha(\omega) = \frac{\sqrt{2}\omega}{c}\left[\sqrt{\epsilon_{1}^{2}(\omega) + \epsilon_{2}^{2}(\omega)} - \epsilon_{1}(\omega)\right]^{1/2}.
\end{equation}
Here $\omega$, is the photon frequency, $ \epsilon_1 = \mathcal{R}[\epsilon (\omega)]$ and $\epsilon_2 = \mathcal{I}[\epsilon (\omega)]$ are the real and imaginary components of the dielectric function, and \textit{c} is the speed of light. The real and imaginary parts of the dielectric function are plotted in Figure S4 in SI, as a function of incident light's energy. As shown in Figure \ref{absorption}, for the in-plane polarized light, the first absorption peak is present in the visible range at $\sim 2.5$ eV for SiH-CdCl$_2$ heterostructure. Following the first peak, several peaks are in the range $\sim 3.5-4.5$ eV in the near-ultraviolet and middle-ultraviolet regions. Calculated absorption coefficients ($\sim 10^6$ cm$^{-1}$) are comparable to perovskite solar cells. A significant absorption coefficient in the visible and near-ultraviolet regions suggests the potential applicability of SiH-CdCl$_2$ heterostructures in photocatalytic and photovoltaic devices.

\begin{figure}
\centering
\includegraphics[width=0.5\textwidth]{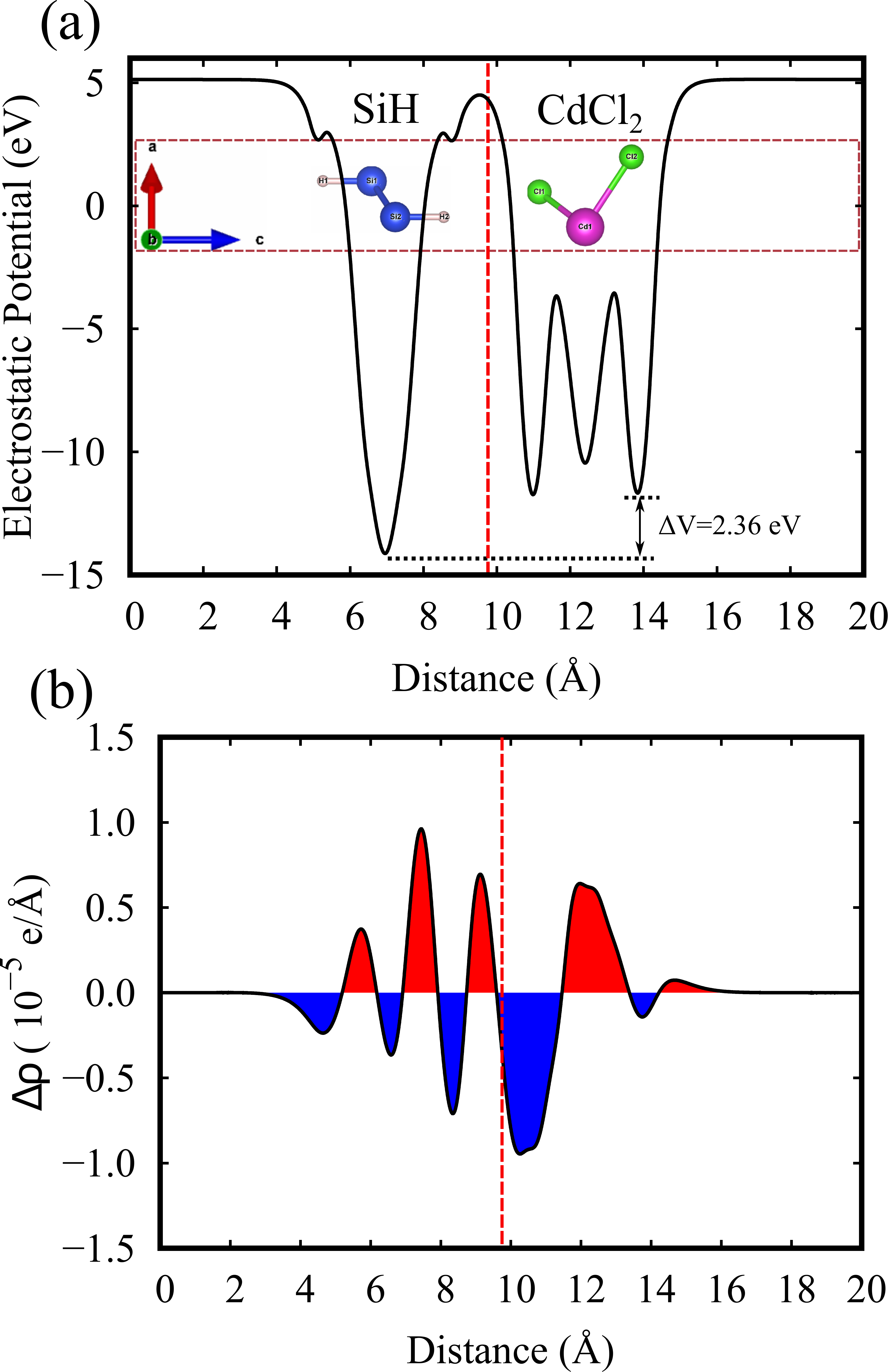}
\caption{(a) Plane averaged potential along the height of the SiH/CdCl$_2$ heterostructure, showing a potential drop of 2.36 eV across the interface. (b) Plane averaged charge density difference ($\Delta \rho$), showing a net accumulation of negative and positive charge in the CdCl$_2$ and SiH side, respectively. The Vertical dashed red line marks the SiH/CdCl$_2$ heterostructure interface as defined in Figure \ref{fig:structure} (c).}
\label{avg_potential}
\end{figure}
\subsection{Heterojunction characteristics}
\label{secef}
Although the constituent monolayers are held together by weak van der Waals interaction, inter-layer charge transfer is still possible because of the electronegativity difference between individual layers. The electrostatic potential is plotted along the $z-$direction [Figure~\ref{avg_potential}(a)], showing a deeper potential at SiH monolayer, compared to CdCl$_2$. The difference between the potential minimum in two constituent monolayers is 2.36 eV, resulting in an internal electric field along negative $z$ (directed from CdCl$_2$ to SiH). As a result, a net charge transfer is expected from the SiH to the CdCl$_2$ monolayer. To confirm this, we plot the planar averaged electron density difference between the heterostructure and its constituent parent compounds. It is defined as, 
\begin{eqnarray}
    \Delta \rho(z)=\int dxdy \left[\rho_{SiH/CdCl_2} -\rho_{SiH} -\rho_{CdCl_2} \right],
\end{eqnarray}
where $\rho_{SiH/CdCl_2}(x,y,z), ~\rho_{SiH}(x,y,z)$, and $\rho_{CdCl_2}(x,y,z)$ are the charge densities of the SiH/CdCl$_2$ heterostructure and the isolated SiH and CdCl$_2$ monolayers, respectively. As shown in Figure~\ref{avg_potential} (b), we find that $\Delta\rho(z)$ is oscillatory, with alternating regions of charge accumulation (red) and depletion (blue). However, a net accumulation of negative (positive) charge can be clearly seen in the CdCl$_2$ (SiH) side, as expected due to the internal electric field.

Because of its unique type-II electronic band structure, optical absorption peaks in the visible and near-ultraviolet range, and internal electric field, the SiH-CdCl$_2$ heterostructure has the potential to be used in several device applications. We discuss some of these applications in the following sections.
\section{Application: Piezoelectricity}
\label{PZE}
\begin{figure}
\centering
\includegraphics[width=0.5\textwidth]{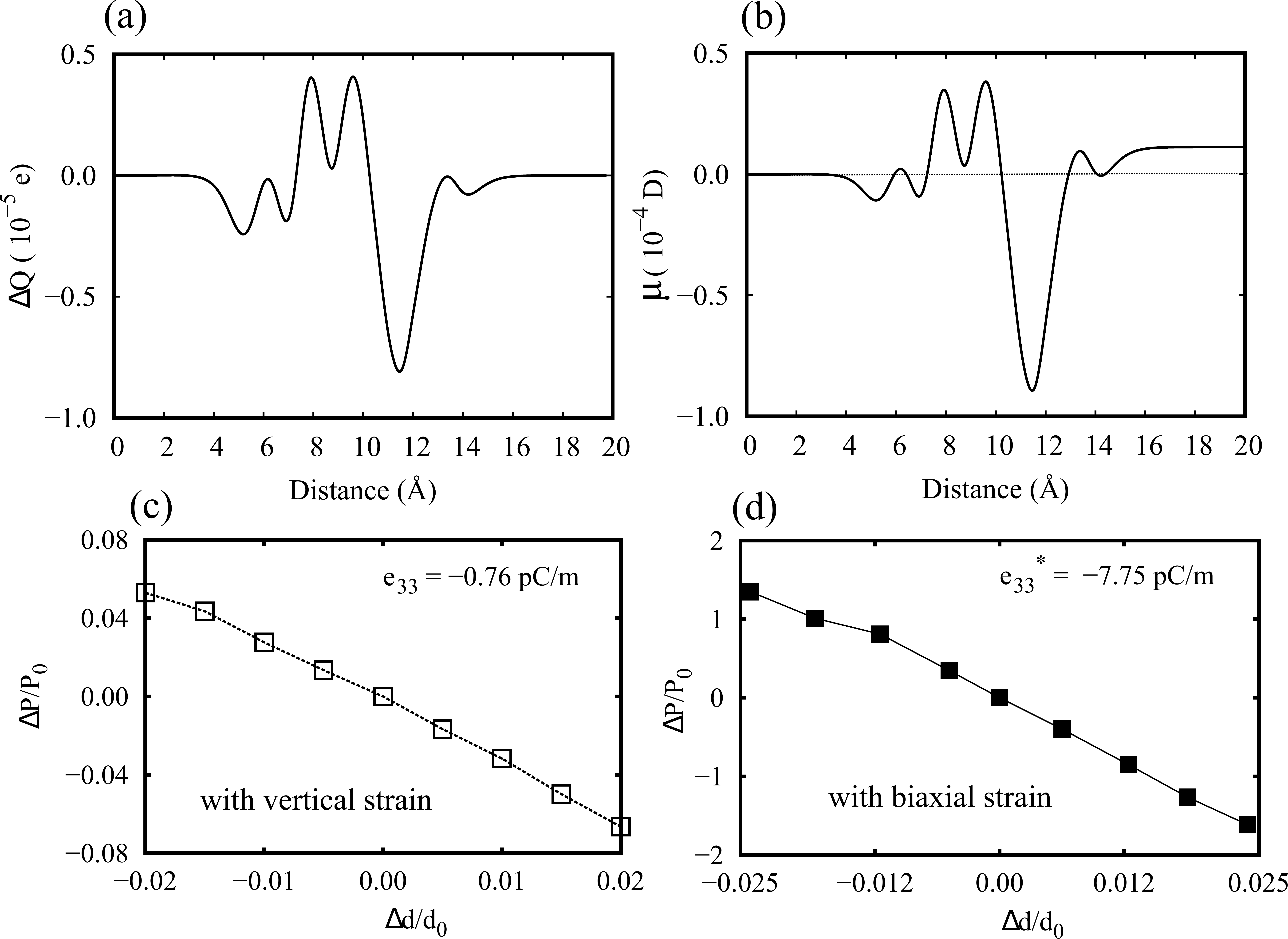}
\caption{(a) Plane averaged value of charge transfer ($\Delta Q$), and (b) dipole moment ($\mu$) in the direction normal to the plane of the SiH-CdCl$_2$ heterostructure. (c)-(d) The change in the out-of-plane polarization as a function of the vertical and biaxial strain, respectively.}
\label{piezoelectric}
\end{figure}
Owing to the built-in electric field, which causes a charge redistribution across the interface [Figure~\ref{avg_potential}(b)], we expect a net electric dipole moment along the out-of-the-plane direction. With the heterostructure configuration shown in Fig.~\ref{fig:structure}(a), the charge transfer from SiH to CdCl$_2$ can be estimated as,
\begin{equation}
\Delta Q(z)=\int_{-\infty}^{z} \Delta \rho (z') \,dz',
\end{equation}
where $\Delta \rho$ is the planar average charge density difference for the SiH-CdCl$_2$ heterostructure [Figure~\ref{avg_potential}(b)]. The plot of $\Delta Q$ [Figure~\ref{piezoelectric}(a)] confirms that the CdCl$_2$ monolayer gains electrons (negative charges) from the SiH monolayer. Using $\Delta Q$ values, interface-induced dipole moment $\mu$ can be calculated as,
\begin{equation}
\mu(z) = \int_{-\infty}^{z} z'\Delta \rho (z') \,dz'.    
\end{equation}
We find the value of $\mu(z)$ to be $+0.13\times 10^{-4}$ debye [see Figure~\ref{piezoelectric} (b)], induced due to the formation of vdW heterostructure of SiH and CdCl$_2$ monolayers. The dipole points from the negatively charged CdCl$_2$ layer to the positively charged SiH layer. The existence of an interfacial dipole moment in the $z$-direction will undoubtedly result in an out-of-plane polarization and hence an out-of-plane piezoelectric coefficient, which we will discuss next.
 
The coefficient of piezoelectricity is described as follows\cite{chen2021enhanced}: 
\begin{equation}
    e_{ijk}=\dfrac{\partial P_i}{\partial \epsilon_{jk}}~.
\end{equation}
Here, $P_i$ is the polarization, and $\epsilon_{jk}$ represents the strain tensor. The in-plane $x$, $y$, and out-of-plane $z$ directions are represented by the values 1, 2, and 3, respectively. Taking polarization along the out-of-plane direction ($i=3$) and longitudinal strain along the out-of-plane direction ($j=k=3$), we calculate the out-of-plane piezoelectric coefficient $e_{333} = e_{33}$ as follows:
\begin{equation}
    e_{33}=\dfrac{\partial P_3}{\partial \epsilon_3}\approx\dfrac{\Delta P_3}{\Delta \epsilon_3}.
\end{equation}
Out-of-plane strain leads to a change in interlayer distance, which can be achieved in two different ways by the application of vertical strain and biaxial strain. If we take $\Delta \epsilon_3\sim \Delta d/d_0$, where $\Delta d$ is the change of interlayer distance from its equilibrium value $d_0$, the change of polarization is much greater in the case of biaxial strain. As a result, $e_{33}$ is one order of magnitude higher [Figures \ref{piezoelectric}(c) and  \ref{piezoelectric}(d)] in the case of a biaxial strain (-7.75~pC/m) than that of a vertical strain (-0.76~pC/m). To understand why biaxial strain has a stronger influence, we compare the change of the intrinsic interlayer potential in the heterostructure at finite strain [Figure S5 in SI]. Clearly, the modulation is much stronger in case of biaxial strain. While applying a perpendicular tensile strain, we only change the vertical distance between the monolayers, keeping intralayer bonds unchanged. On the other hand, biaxial strain not only changes the interlayer distance, but also modifies the intralayer bonds in individual monolayers, leading to a stronger effect on polarization. Interestingly, we find that the sign of the piezoelectric coefficients is negative, which makes this type-II heterostructure suitable for many device applications, such as flexible transducers and energy harvesters \cite{urbanaviciute2019negative}.  Other van der Waals materials like polyvinylidene fluoride, CuInP$_2$S$_6$, and BiTeX (X = Cl, Br, and I), are also reported to have a negative piezoelectric response \cite{katsouras2016negative, PhysRevB.100.104115, liu2016room}. Next we explore the photocatalysis potential of SiH-CdCl$_2$ heterostructure. 

\section{Application: photocatalysis}
\label{PhotoCt}
\begin{figure}
\centering
\includegraphics[width=0.45\textwidth]{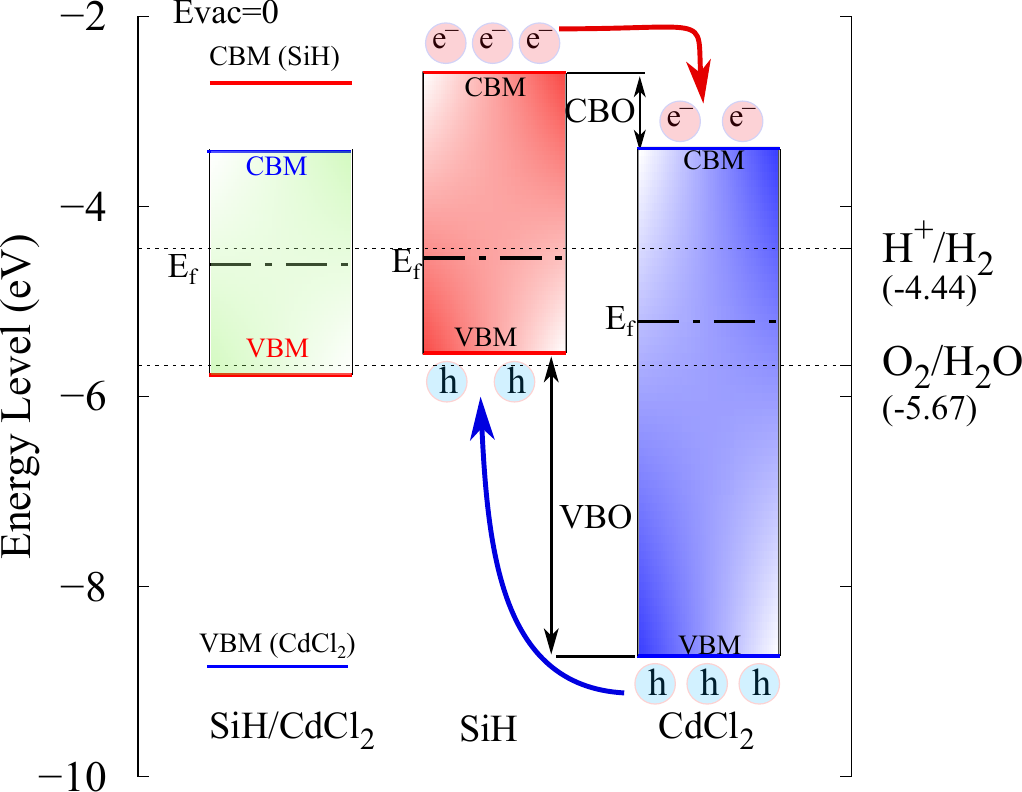}
\caption{Band diagram of isolated monolayers of SiH, CdCl$_2$, and their heterostructure. The plot also shows the energy levels for oxidation and reduction potential for reaction related to water splitting with respect to vacuum level (0 eV). We also mark the valence band offset (VBO) and the conduction band offset (CBO) as they are responsible for the separation of charge carriers produced by photons.}
\label{band_diagram}
\end{figure}
Photocatalytic water splitting requires water to be simultaneously oxidized (2H$_2$O + 4h$^+$ = 4H$^+$ + O$_2$) and reduced (4H$^+$ + 4e$^-$=2H$_2$), such that the overall result is 2H$_2$O = 2H$_2$ + O$_2$. The holes (h$^+$) and electrons (e$^-$) required for the process are generated by the absorption of photons. Because of the type-II band alignment, the oxidation reaction occurs at the heterostructure's SiH layer (VBM), while the reduction reaction occurs at the CdCl$_2$ layer (CBM). A large absorption peak present at $\sim 2.5$ eV [Figure~\ref{absorption}] 
ensures the availability of abundant photogenerated charge carriers in the presence of solar light. The physical separation of the oxidation (induced by holes) and reduction process (induced by electrons) ensures higher efficiency of the photocatalytic process. 

Additionally, the type-II band alignment of SiH-CdCl$_2$ heterostructure has other advantages for photocatalytic applications. The band diagram of SiH-CdCl$_2$ heterostructure is illustrated in Figure~\ref{band_diagram}. The band edges are calculated with respect to the vacuum level. Band edges (VBM and CBM) of monolayer SiH are at higher energy levels than those of the monolayer CdCl$_2$. Even in the heterostructure, respective band edges change marginally. There is a valence band offset (VBO, between VBM of SiH and CdCl$_2$) of 3.1 eV and a conduction band offset (CBO, between CBM of SiH and CdCl$_2$) of 0.8 eV. The band gap between the VBM of SiH and CBM of CdCl$_2$ is 2.43 eV. As shown in Figure~\ref{band_diagram}, because of the CBO and VBO, electrons and holes are accumulated at the CdCl$_2$ and SiH layer, respectively. With charge separation in different layers, electrons and holes will have a low recombination rate, thus offering more efficiency for photocatalysis. A significant difference between electron and hole mobility [see Table~\ref{mobility}] further helps reduce the recombination rate and increase the lifetime of charge carriers. As discussed earlier, owing to low effective mass, hole mobility is almost six times higher than that of electrons. As a result, the photogenerated holes move away faster than the slower electrons, reducing electron-hole recombination and enhancing the efficiency of SiH-CdCl$_2$ heterostructure as a photocatalyst.

In addition to the properties mentioned above, energy levels of reduction (H$^+$/H$_2$) and oxidation (H$_2$O/O$_2$) potential should lie within the band gap of the photocatalyst for water splitting. As shown in Figure~\ref{band_diagram}, the reduction (-4.44 eV) and oxidation (-5.67 eV) potential lie within the bandgap of CdCl$_2$ monolayer, but the VBM of the SiH monolayer is more positive than the (H$_2$O/O$_2$) potential, making the latter unsuitable as a photocatalyst for water splitting. However, the reduction and oxidation potential lies within the bandgap of SiH-CdCl$_2$ heterostructure, making it suitable for photocatalysis. As already explained, SiH-CdCl$_2$ heterostructure is a better photocatalyst than monolayer CdCl$_2$ because the latter has lower efficiency due to electron-hole recombination. The standard oxidation-reduction potentials are a function of pH [see Eq. S1 and S2 of the SI]. The values reported in Figure~\ref{band_diagram} are for a pH=0. We have verified that the oxidation-reduction potentials lie within the bandgap of SiH-CdCl$_2$ heterostructure from pH 0 to 14 [see Figure S6 in SI], ensuring a wide range of applicability for 
photocatalytic application. Having demonstrated the photocatalytic abilities of the SiH-CdCl$_2$ heterostructure, we now explore its potential as a field effect transistor.

\section{Application: tunnel field effect transistor (TFET)}
\label{TFET}
\begin{figure} 
\includegraphics[width=0.9\linewidth]{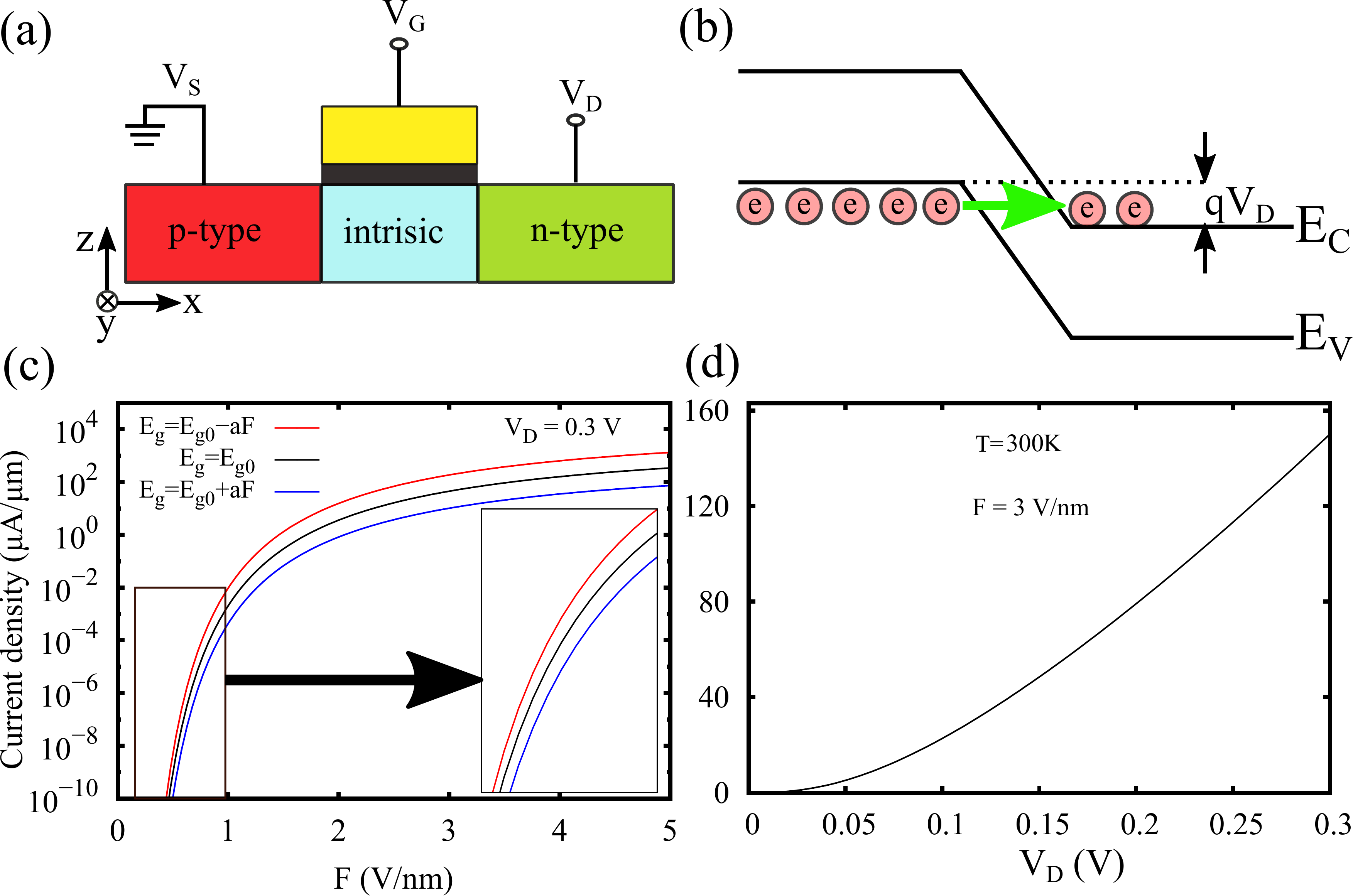}
\caption{ (a) Schematic depiction of a 2D crystal p-i-n junction based TFET, (b) Band diagram of TFET, (c) Interband tunneling current density for \ch{SiH/CdCl2} heterostructure at a bias of V=0.3~V and (d) Current density as a function of voltage.} 
\label{fig:TFET}
\end{figure}
The ability of a heterojunction based vertical two-dimensional TFET to overcome the thermionic limit of subthreshold swing (SS) (60 mV/dec) makes it one of the most promising options for ultra-thin and low-power integrated circuits. The analytical model proposed by Jena \cite{ma2013interband}, suggests that the interband tunneling current  of a vdW-based TFET can be calculated using 
\begin{multline}
    J_{T}^{2D}=\frac{qg_{s}g_{v}T_{0}k_{B}T}{\left(2\pi \right)^{2}\hbar} \int\limits_{-K_{max}/\eta}^{+K_{max}/\eta} dk_{y} ~ exp\left[ -\frac{E_{y}}{\bar{E}} \right]\\
\times \ln \left\lbrace \frac{\left( e^{\beta\left( qV-\eta^{2}E_{y} \right)     }+  e^{-\beta E_{y}} \right)  \left(  1+e^{\beta\left( qV-E_{y}\right) }     \right)  }{\left( e^{\beta \left(qV-\eta^{2}E_{y} \right) }+ e^{\beta\left(qV-E_{y} \right) } \right)\left( 1+e^{-\beta E_{y}}\right)}     \right\rbrace
~.
\label{multline:1}
\end{multline}
Here, WKB tunneling probability $T_0=\exp[-4\sqrt{2m_{R}^{*}}E_{g}^{3/2}/3q\hbar F]$, $\bar{E}=q\hbar F/2\sqrt{2m^{*}_{R}E_{g}}$, $E_{y}=\hbar^2K_{y}^2/2m_{v}^{*}$, and $K_{max}^2=2m^{*}_{v}qV/\hbar^2$. Other terms needed are as follows: $q$ is the charge, $g_{s}$ and $g_{v}$ are spin and valley degeneracy, reduced mass $m^{*}_{R}=m_{c}^{*}m^{*}_{v}/(m_{c}^{*}+m^{*}_{v})$, $m_{c}^{*}$ and $m^{*}_{v}$ are the effective masses of the valence and conduction band extremum, and $F$ is the vertical field. Further, $\eta$ and $\beta$ are defined as  $\eta^2=m^{*}_{v}/m^{*}_{R}$ and $\beta=1/k_{B}T$, where $\hbar$ and $k_{B}$ are the reduced Plank constant and Boltzmann constant, respectively. For SiH/CdCl$_2$ heterostructure, the electronic band gap $E_{g}$ changes linearly with the applied field [see Figure~\ref{fig:E-Eg}(b)]. We can express it as $E_{g}=E_{g0}\pm aF$, where $E_{g0}$ is the zero-field band gap of the SiH/CdCl$_2$ heterostructure. The constant $a=1.56$ is calculated from a linear fit of the $E_{g}$ vs. field plot in Figure \ref{fig:E-Eg} (b). We also verify that linearity holds good for very small electric field values [see Figure S7 in SI]. 

The calculated tunneling current density as a function of vertical field and voltage (at different temperatures) are shown in Figures \ref{fig:TFET}(c) and \ref{fig:TFET} (d), respectively. We find that the interband tunneling happens only when $V$ exceeds zero since the zero drain voltage $V=0$ is defined when the source's conduction band edge is aligned with the drain's valence band edge. Due to the exponential dependence of the tunneling probability on the field $F$ ($T_{0}\propto exp[-4\sqrt{2m_{R}^{*}}E_{g}^{3/2}/3q\hbar F]$), the tunneling current varies exponentially for low values of F [Figure \ref{fig:TFET}(c)], and begins to saturate when $F$ exceeds 1.5 V/nm. The tunneling transistors based on 2D transition metal dichalcogenides (TMDs) show a similar trend \cite{ma2013interband, jiang2014performance, balaji2018tunneling, pal2020compact}.

The change of bandgap ($E_g$) with external field [Figure~\ref{fig:E-Eg}] also affects the current, as the tunneling probability increases (decreases) due to a decrease (increase) in $E_g$. We carry out a systematic investigation by calculating the current in three different cases: I) $E_g$ decreases with $F$, II) $E_g$ remains unchanged, and III) $E_g$ increases with $F$. While case I and III are observed in SiH/CdCl$_2$ [Figure~\ref{fig:E-Eg}], case II is calculated to fully understand the extent of the effect of bandgap tuning via an external electric field. As expected, the lower the bandgap, the higher the tunneling current, which is the highest in case I, followed by case II, and lowest in case III. As shown in Figure \ref{fig:TFET}(c), the on-state current = 1300$~\mu A/\mu m$ for case I is about 4 and 18 times greater than that in case II and case III, respectively. Note that case I also has a better SS [Figure~\ref{fig:TFET}(c)] than the rest of the cases. In all cases, SS is less than 60 mV/dec, with the lowest SS in case I.


The tunneling current as a function of voltage at the room temperature is presented in Figure~\ref{fig:TFET}(d). In contrast to typical metal-oxide-semiconductor field-effect transistors (MOSFETs), we observe a rectifying curvature. Several studies conducted on TFETs have demonstrated a similar trend \cite{mallik2011drain, pal2011insights, lattanzio2011complementary}. Such a quasi-exponential behavior of current characteristics is typical of band-to-band tunneling (BTBT) and is due to the drain barrier modulation of the BTBT window. As depicted in Figure~\ref{fig:TFET}(d), at small reverse bias voltages $V\ll V_{0}=\eta^2 \bar{E}/q $, the tunneling current changes as $J^{2D}_T\sim V^{1.5}$ \cite{ma2013interband}. For larger values of $V\gg V_{0}$, Eq.~\ref{multline:1} changes to $J^{2D}_T=G\times V$, where $G\approx [q^2/h(g_sg_v/2\pi)\sqrt{(2\pi m^{*}_v/\hbar^2)\times (q\hbar F/\sqrt{8m^{*}_R E_g})}\times T_0]$ and the current thus varies linearly with V \cite{ma2013interband}.


\section{Conclusion}
\label{conclusion}
Using \textit{ab initio} calculations, we demonstrate the potential of a newly predicted heterostructure formed by SiH and CdCl$_2$ monolayers for different device applications. The heterostructure has a direct bandgap of 2.43 eV and a Type-II band alignment in which the VBM and CBM are localized to the SiH and CdCl$_2$ monolayer, respectively. Such a band alignment is beneficial for photocatalytic water splitting, as electrons and holes are separated at two different monolayers, reducing their recombination rate and increasing device efficiency. We show that the optical absorption spectrum has peaks in the visible wavelength, confirming photocatalytic device operation using solar energy. Our calculations further reveal that the heterostructure can be explored for making TFET devices. 

Our study highlights that the characteristics of the heterostructure mainly depend on the potential difference between the constituent monolayers, leading to the charge transfer, which is the origin of piezoelectricity. With many theoretically predicted and experimentally realized layered materials,  we can form many combinations of heterostructures with a suitable potential difference to engineer specific  functionalities. Looking at the potential for various functionality, we believe that an exciting future lies ahead for device applications in assembled heterostructures. 

\section*{ACKNOWLEDGMENTS}
We acknowledge National Super Computing Mission (NSM) for providing computing resources of “PARAM Sanganak” at IIT Kanpur, which is implemented by C-DAC and supported by the Ministry of Electronics and Information Technology (MeitY) and Department of Science and Technology (DST), Government of India. We also acknowledge the HPC facility provided by CC, IIT Kanpur. We acknowledge the Science and Engineering Research Board (SERB, file no: EMR/2017/004970) for financial support.



\bibliography{references}

\end{document}